# Atomically resolved probe-type scanning tunneling microscope for use in harsh vibrational cryogen-free superconducting magnet

Wenjie Meng[1,2#], Jihao Wang[1,2#], Yubin Hou[1,2*], Mengqiao Sui[3], Junting Wang[4], Gang Wu[3], Junyun Li[3] and Qingyou Lu[1,2,4,5*]

*1. Anhui Province Key Laboratory of Condensed Matter Physics at Extreme Conditions, High Magnetic Field Laboratory of the Chinese Academy of Sciences, Hefei, Anhui 230031, People's Republic of China*

*2. CASmF Sci.&Tech. Ltd.（合肥中科微力科技有限公司） Hefei, Anhui 230088, People's Republic of China*

*3. NanoScience devision, Oxford Instruments Technology (Shanghai) Co., Ltd,. Shanghai, 200233, People's Republic of China*

*4. Hefei National Laboratory for Physical Sciences at the Microscale, University of Science and Technology of China, Hefei, Anhui 230026, People's Republic of China*

*5. Collaborative Innovation Center for Artificial Microstructure and Quantum Control, Nanjing University, Nanjing 210093, People's Republic of China*

*Correspondence should be addressed: ybhou@hmfl.ac.cn and qxl@ustc.edu.cn*

**Abstract**

We present a probe-type scanning tunneling microscope (STM) with atomic resolution that is designed to be directly inserted and work in a harsh vibrational cryogen-free superconducting magnet system. When a commercial variable temperature insert (VTI) is installed in the magnet and the STM is in turn housed in the VTI, a lowest temperature of 1.6 K can be achieved, where the STM still operates well. We have tested it in an 8 T superconducting magnet cooled with the pulse-tube cryocooler (PTC) and obtained atomically revolved graphite and $NiSe_2$ images as well as the scanning tunneling spectrum (STS, i.e. dI/dV spectrum) data of the latter near its critical temperature, which show the formation process of the superconducting gap as a function of temperature. The drifting rates of the STM at 1.6 K in X-Y plane and Z direction are 1.15 and 1.71 pm/min respectively. Noise analysis for the tunneling current shows that the amplitudes of the dominant peaks (6.84 and 10.25Hz) are low. This is important as a cryogen-free magnet system has long been considered too harsh for any atomic resolution measurement.

**Key words:** probe-type scanning tunneling microscope; cryogen-free superconducting magnet; cryogen-free magnet; scanning tunneling spectroscopy; pulse-tube or GM cryocooler.

## 1. Introduction

Scanning tunneling microscopy (STM) including scanning tunneling spectroscopy (STS) has become the most celebrated technique as it is capable of providing not only the atomically resolved local electronic density of states (LDOS), but also the structure of energy band with high resolution [1]. Combined with low temperature and high magnetic field conditions, the STM plays a unique and valuable role in acquiring atomic scale information from both real and momentum spaces in condensed matter physics [2]. Such STM facility capable of figuring out complete H-T phase diagram is particularly popular [3, 4], which helps us access to exotic electronic phases and quantum effects, such as the symmetry of the superconducting order parameter in a superconducting material from analyzing quasi-particle interference [5, 6], Fermi arc of a Weyl semimetal [7], quantum nano-structure on the surface of a semiconductor, pseudo-gap state in a high temperature superconductor [8] and so on.

However, the STM has a major drawback: extremely sensitive to external vibrations, even including normal acoustic noise [9]. Hence, a high field STM has to be housed in a wet superconducting magnet, where the superconducting coil is immersed in liquid helium in a bath cryostat, thus having the superiority of tranquility [3, 10]. But now comes a severe issue: although a liquid helium bath cryostat generates weaker vibration, it consumes large amount of liquid helium. The cost will soar in the near future since helium is a limited and non-renewable resource. As a matter fact, we are now experiencing a worldwide helium shortage and predictable worse situation for basic research [11, 12]. Apart from the high liquid helium consumption, the sophisticated design of cryostat required to maintain long holding time and intense labor investment for running the system add huge cost for the researchers. The solution is well known: using a cryogen-free magnet system, which does not consume any liquid helium as the superconducting coil is cooled by a PTC or GM cryocooler and easy for operation. From the rapid growing applications of cryogen-free magnet system, it is very clear that they represent the trend of future.

Now the question comes back again: the vibration condition for a cryogen-free magnet system has long been considered way too harsh for any atomically resolved STM measurement. This is indeed a fundamental issue we need to solve very seriously, which holds not only for cryogen-free magnet but also for many other application situations. To this end, we have been working on the implementation of atomic resolution STM under harsh conditions, which has finally led to the successful construction of the world's first STM in an ultra-harsh water-cooled magnet [13], where atomically resolved STM images were obtained in a magnetic field up to 27 T, stronger than the field produced by any superconducting magnet. In this paper, we will present a probe-type STM with atomic resolution which works by conveniently inserting it in a cryogen-free superconducting magnet system.

Our STM probe has been proved working well at temperature down to 1.6K and in magnetic field up to 8T in a cryogen-free magnet system. We obtained atomically revolved graphite and $NiSe_2$ images as well as the scanning tunneling spectrum (STS, i.e. dI/dV spectrum) curves of $NiSe_2$ near its transition temperature $T_C$, which reveal the dynamic formation process of the superconducting gap as the temperature drops from above $T_C$ to below $T_C$. The noise spectrum for the tunneling current shows that the amplitudes of the dominant peaks are fairly low. As an obvious advantage of the almost non maintenance required cryogen-free magnet system, both the STM and the magnet can operate remotely through the internet for a preloaded sample for weeks without on-site operation.

Previously, only very few groups can be successful in obtaining atomically resolved STM images on a cryogen-free superconducting magnet system and they achieve that mainly in a complicated way in which the STM is hung in the harsh vibration magnet (without touching it) through an external table with vibration isolation [14]. Such an externally supported STM system has a few severe drawbacks. In addition to the high cost from the external support with complicated vibration isolation setup, the STM has to be bound to the expensive magnet permanently since removing the STM from the magnet is very inconvenient. This is a big waste of resource as no other equipment can access to the magnet any longer. The importance of the plug-and-play convenience of our probe-type STM is hence easily seen and no similar achievement has been published before. The compact design of the probe-type STM makes it easily compatible with any other cryogen-free platform in principle.

## 2. System design

### 2.1 Cooling system

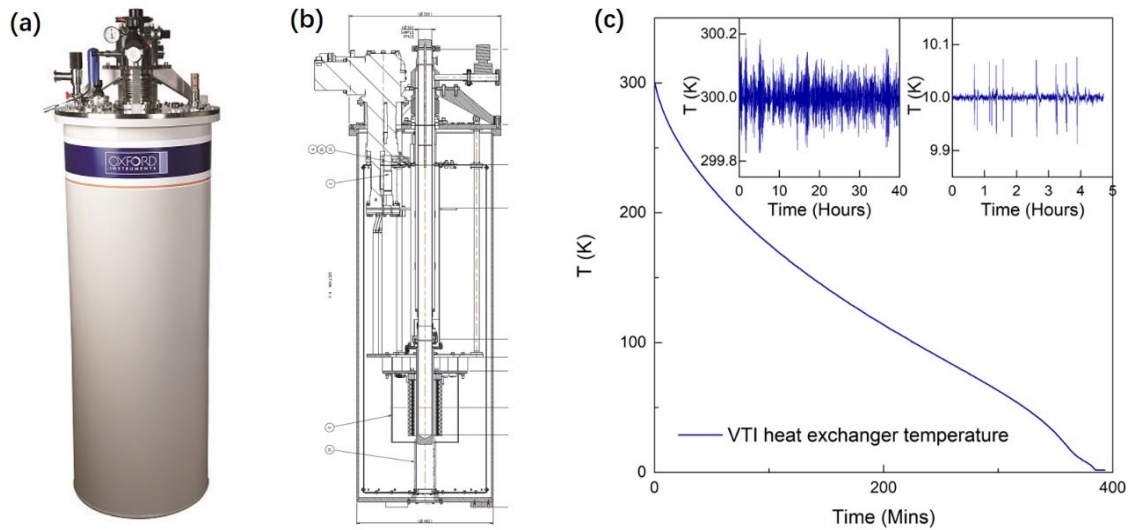

Fig. 1(a) and Fig. 1(b) are photograph and cross section drawing of the 8T Cryofree superconducting magnet system. Fig.1(c) shows typical cool down curve of VTI heat exchanger from room temperature to base temperature. Typical temperature stability over long period at 300K and 10K are presented in the insets.

The experiments were performed in Oxford Instruments Shanghai demo lab with a cryogen-free superconducting magnet system (model: TeslatronPT from Oxford Instruments, Fig1(a) ), which provides top loading access to a sample in a variable magnetic field / low temperature environment. The system used in this paper provides magnetic field up to 8 T, with an integrated VTI providing sample temperature from 1.6 K to 300 K and isolated cylindrical sample space with diameter of 50mm, Fig1(b). When the STM probe was inserted into the sample space, the sample and probe are in helium gas atmosphere and cooled down by the static exchange helium gas. Typical cool down time is about 400mins, Fig1(c). VTI temperature is monitored and controlled with a PID feedback loop. VTI temperature sensor and heater locates at the VTI heat exchanger, which is 10cm above magnet top plate on the VTI outer shield. Typical temperature variation is +/- 50mK during all the experiments. Temperature stability curves at 300K and 10K are shown in Fig1(c) insets. 8T superconducting magnet is located at the bottom of the cryostat. When the STM probe sits inside the VTI, the sample stays at the magnet center position. An estimated absolute vibration level at the sample position is less than +/- 10um in horizontal direction.

**2.2 Design of probe**

A photograph of the STM probe is shown in Fig.2(a). The probe mainly consists of pre-amplifier circuit box of outer diameter 75 mm and 39 mm height, bellow joint of outer diameter 25 mm, central tube of outer diameter 10 mm, three radiation shield plates of diameter 46 mm, heavy rod of outer diameter 35 mm made of Teflon and brass, sapphire interface piece of outer diameter 30 mm with multiple pins, vibration isolating spring and STM head. The pre-amplifier circuit is placed in the circuit box, which is positioned at the top end of the central tube and connected and vacuum sealed with the central tube via the bellow joint. The circuit box is the only part that is outside the magnet. All the remaining parts form the inner part of the probe, which is housed in the magnet. The soft bellow joint can at least partially prevent the vibration of the cryogen-free magnet system from entering into the inner part of the probe. At the bottom of the pre-amplifier circuit box is a standard KF50 flange, which can vacuum seal with the flange provided at the entrance of the magnet.

Three radiation shield plates are made from 304 stainless steel. For each plate, both sides are polished in order to reduce the heat exchange between the STM head at low temperature and the external environment at room temperature. They are coaxially mounted on the central tube. The lower end of the central tube is mounted with the heavy rod such that the central tube is coaxially connected with the heavy rod in series with a long through hole at the center for passing wires. The heavy rod is designed to decrease the resonant frequency of the probe, thus enhancing the vibration isolation capability. It consists of three Teflon and three brass weights that are alternately mounted in series. This arrangement can help reduce the heat exchange between its top and bottom ends. This reduction of heat exchange is important as the heavy rod is thick and long and can easily cause severe thermal short in the cold bore of the magnet, which is normally filled with some low pressure helium gas for thermal exchange purpose.

The sapphire interface piece is directly mounted underneath the heavy rod. The STM head is hung underneath the heavy rod via the vibration isolating spring. Here, owing to gravity, the heavy rod can also ensure that the STM head is hung at the center of the magnet bore. The measurement and control signal wires from the STM head all go upward and are plugged onto the sapphire interface piece from bottom. From the top side of the sapphire interface piece, new wires go upward, pass through the central hole of the heavy rod and central tube and are connected to the vacuum sealed interface (sapphire made) at the bottom of the pre-amplifier circuit box.

**2.3 STM head**

A schematic drawing of the STM head in section view is shown in Fig.2(b). It mainly consists of five coaxially installed parts as follows: (1) tubular main frame (tantalum made) of 19 mm outer diameter, (2) piezoelectric tube (0.5 mm thick EBL#3 type material from EBL Products Inc.) driven inertial motor, (3) four-quadrant piezoelectric tube (0.5 mm thick EBL#3 type material from EBL Products Inc.) scanner, (4) tip holder (sapphire made), and (5) sample holder (tantalum made).

For the piezoelectric tube driven inertial motor, we have actually introduced a similar one elsewhere [15]. It operates by using a tubular piezoelectric actuator of 10 mm outer diameter and 40 mm length to repeatedly move a square tantalum shaft in a stick-slip manner, which is spring-clamped inside the free end of the actuator. In other words, it is essentially an inertial motor of stick-slip type. The purpose of choosing this design is its compactness, which is important for the applications in a small bore magnet with strong vibrational disturbance. Other designs for the motor with high compactness should also work, such as the ones introduced in Ref. [16].

The inertial motor is fixed inside one end of the main frame. In front of the inertial motor, the piezoelectric tube scanner of 3.2 outer diameter and 7 mm length is spring-clamed inside the main frame in a manner similar to that described in Ref. [17], which can be axially pushed to move by the shaft of the inertial motor for the implementation of coarse approach [18]. The tip holder is mounted at the free end of the piezoelectric tube scanner, which approaches the sample holder (mounted inside the other end of the main frame) under coarse approach. The STM head is designed to be hung underneath the heavy rod via a long and soft spring for vibration isolation as described above. In such a design, the coarse approach is actually done downward, which becomes much easier under the impact of gravity.

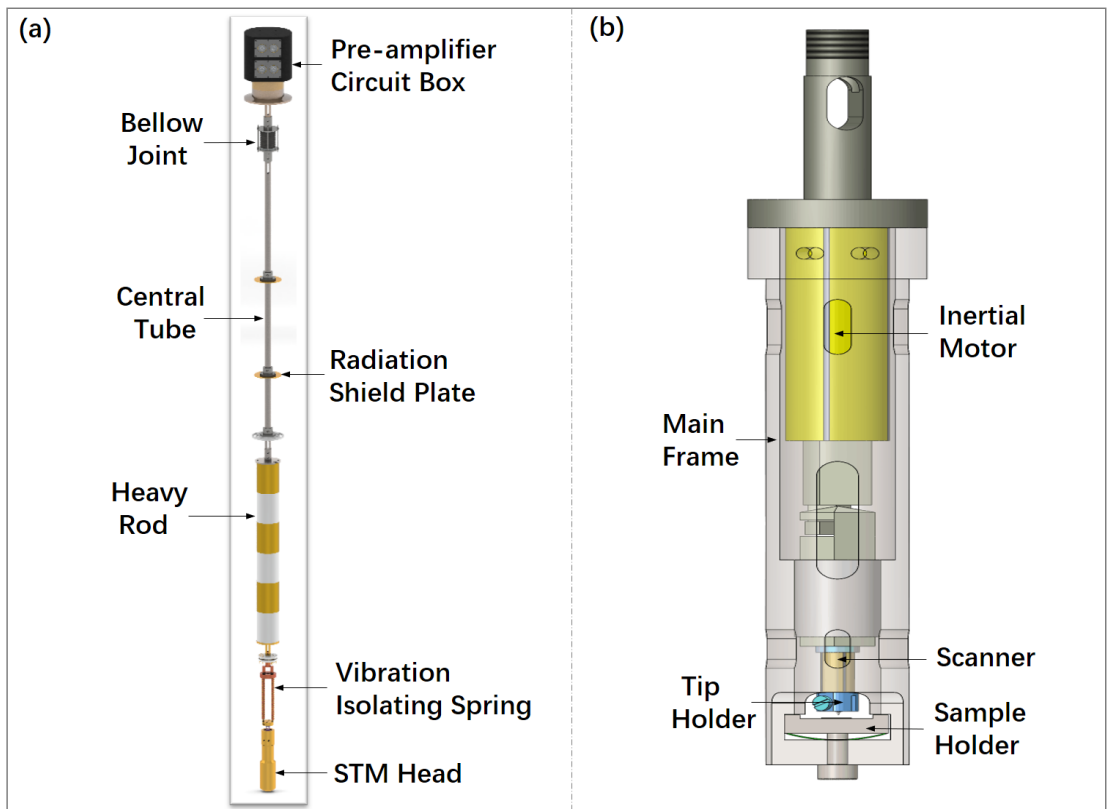

Fig.2. Photograph of the probe-type STM and the section view of the STM head, in which the main parts are indicated.

### 2.4 Control system

We use a low-noise controller (product of CASmF Sci. & Tech. Ltd.) for the probe-type STM to perform the coarse approach and image scan. Basically, the electronics of the control unit is built using a PXI Chassis (PXI-1082Q) from National Instruments10 that is equipped with an embedded controller (PXI-8108) and a data acquisition card of model PXI-7841R. The controller features eight low-voltage analog outputs and eight analog inputs implemented using field-programmable gate array (FPGA) technology to provide the necessary electrical signals, and five independently controlled high-voltage analog outputs (200V maximum voltage). The high-voltage signals are used to control the coarse approach inertial motor as well as the Z adjustment of the piezoelectric scanner in the STM head. The low-voltage output signals are used to control the X-Y scans of the scanner, as well as to apply a +200 mV bias voltage on tip. The low voltage input signals are used to acquire tunneling current and $dI/dV$ spectroscopy data. The control unit is addressed over a transmission control protocol/Internet protocol (TCP/IP) connection using a LABVIEW-based software that runs on computer. Under such a design scheme, the STM can be operated remotely via Internet.

## 3. Performance test

Since the noise from PTC is mainly concentrated toward low-frequency end (1.4Hz and higher orders), we choose to use an 125 Hz low pass filter to define the bandwidth of the preamplifier, which is enough to calibrate the external influence. In Fig.3, we show the measured tunneling current spectrum data for different tunneling current values under constant height mode. The amplitudes of the dominant peaks (at 6.84 and 10.25 Hz) are low enough compared with the STM made by RHK Technology, Inc. [19] for STS measurements.

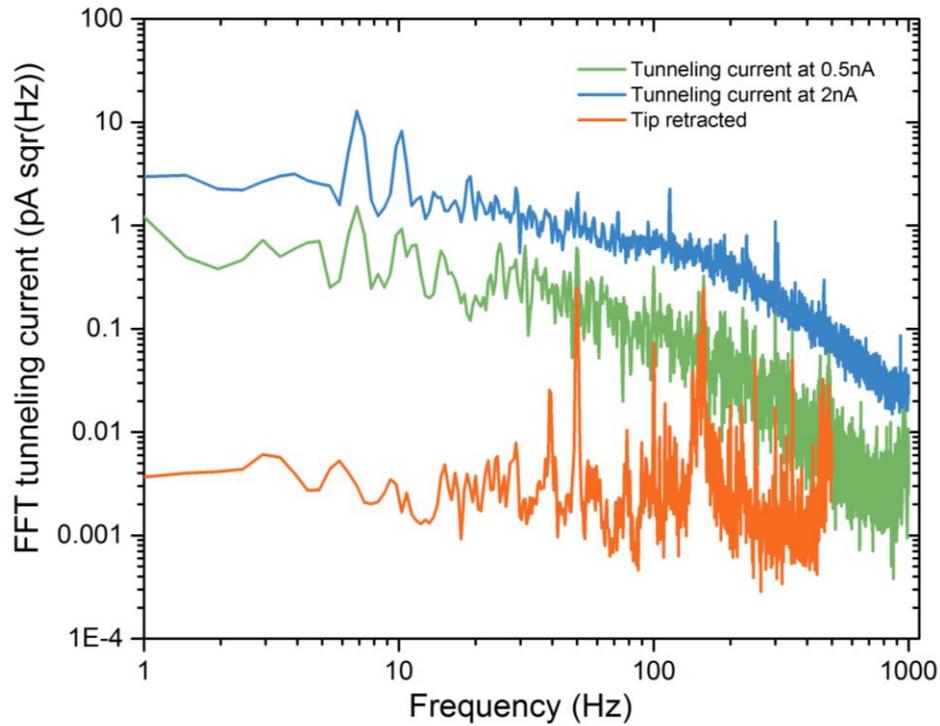

Fig.3. Tunneling current spectrum data at different tunneling current values. The data were taken in constant height mode with a +200 mv bias voltage being applied on NbSe2 sample. The sample was cleaved under ambient condition. The orange curve is the tunneling current noise obtained when the tip was outside of tunneling range and serves as reference.

Fig.4 shows the imaging performance of the probe-type STM system in the cryogen-free magnet. The atomically resolved STM image of graphite (Fig.4(a)) was taken at 50 K in zero magnetic field but with the pulse-tube cryocooler running. Fig.4(b) is the tunneling current line cut profile curve measured along the marked green dashed line in Fig.4(a), which shows high smoothness with excellent signal-to-noise ratio. To test the time stability of the probe-type STM when the PTC is ON, we present the drifting rate measurements in X-Y plane and Z direction at 1.6K in Fig.4(c), which correspond to 1.15 and 1.71 pm/min, respectively, based on the slopes of the drifted distance vs. time curves.

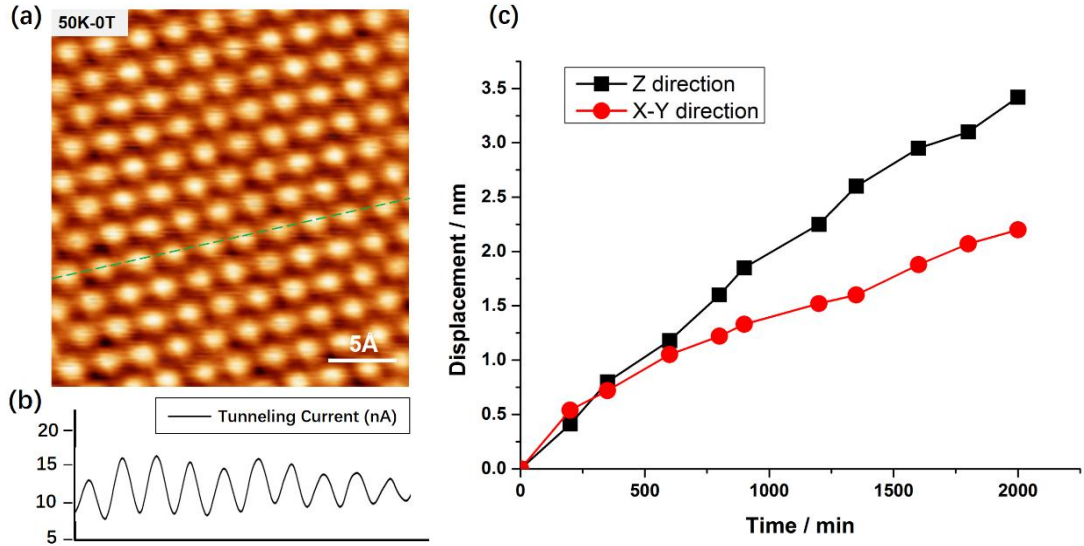

Fig.4. (a) STM image of graphite taken under constant height mode at 50 K in a cryogen-free superconducting magnet. Scan area = 2.3 nm ×2.3 nm, bias voltage = + 200 mV (Tip positive), and average tunneling current =10 nA. (b) Cross-sectional profile for the line cut marked by the green dashed line. The obtained curve is fairly smooth. (c) Measurements of X-Y and Z drifting rates at 1.6 K in the cryogen-free magnet system with the PTC ON, which gives 1.15 and 1.71 pm/min, respectively, according to the slopes of the drifted displacement vs. time. The data were collected after the system was stabilized for an hour.

Using our probe-type STM, we were able to scan a series of images in succession for the graphite sample at 50 K in varying magnetic fields from 0 - 8 T (see Fig.5), which shows high immunity to the external vibration and changing magnetic field. Finally, as an example of a practical application of the cryogen-free magnet system based probe-type STM, we used it to investigate a $NiSe_2$ single crystal sample which had a superconductive transition temperature $T_C$ near 7 K. Fig.6(a) - (d) are the measured STM images at (300 K, 0 T), (300 K, 8 T), (1.6 K, 0 T) and (1.6 K, 8 T), respectively. Fig.6(e) shows the dI/dV spectra of the $NiSe_2$ sample obtained at different temperatures ranging from 1.6 to 8 K, which reveal the dynamic process of energy gap formation when the temperature drops from above $T_C$ to below $T_C$. Above 7 K, no energy gap is seen, whereas below 7 K, the energy gap appears and which gradually becomes smaller as the temperature goes down. These are consistent with theory.

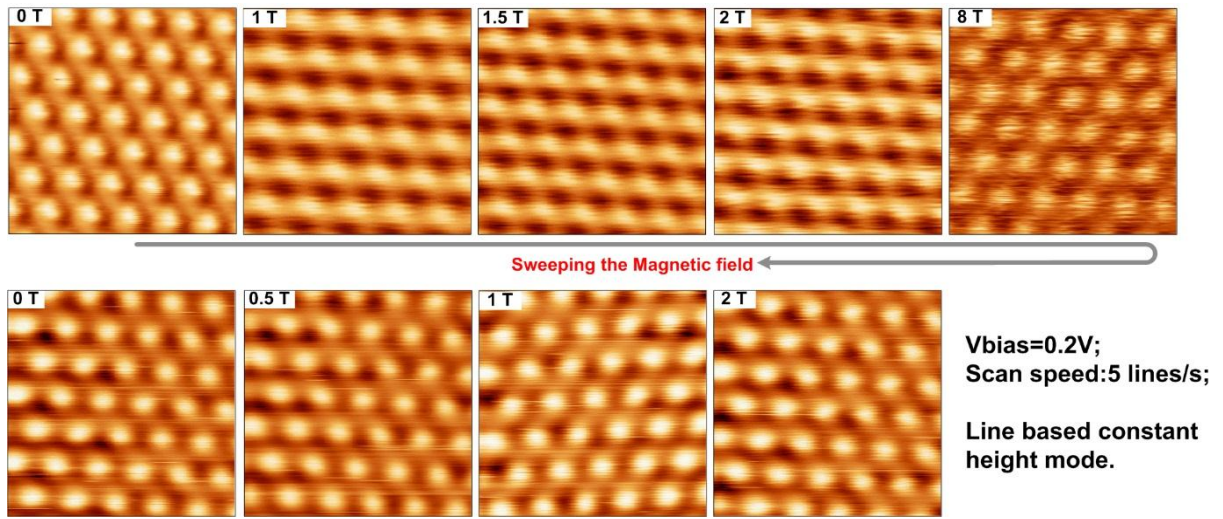

Fig.5 Series of STM images of graphite obtained in sweeping magnetic field from 0 -8 T at 50 K. Scan area = 1.5 nm × 1.5 nm, average tunneling current = 5 nA.

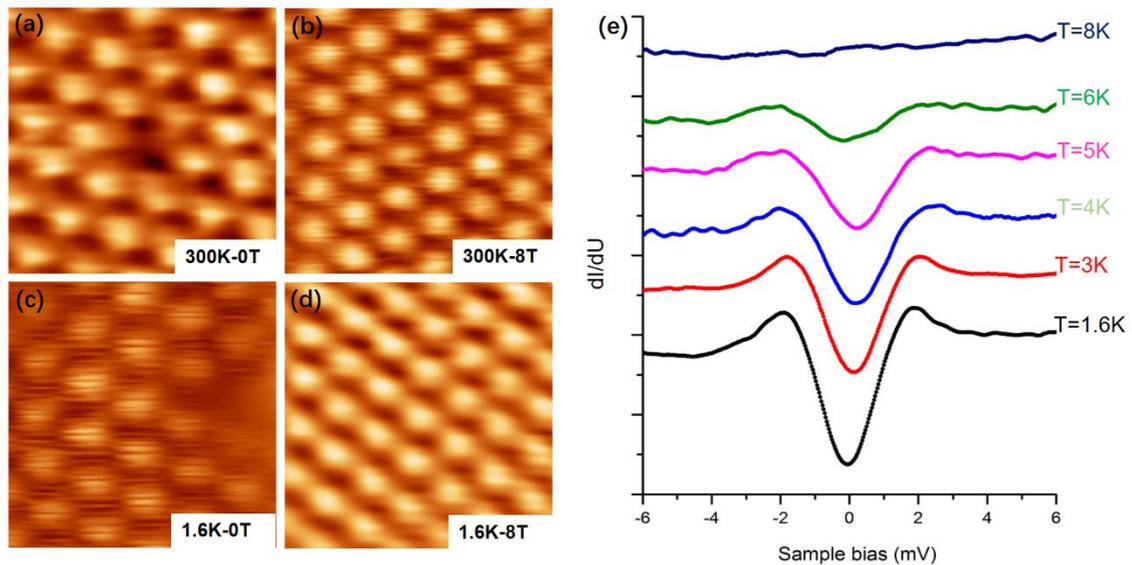

Fig.6 STM images and dI/dV spectra of NiSe2. Scan area = 1.5 nm × 1.5 nm and average tunneling current = 5 nA.

## 4. Conclusion

We have present the design and performance of a probe-type STM operated in an 8 T high field and 1.6 K low temperature cryogen-free superconducting magnet. The tests for atomically resolved imaging, noise spectrum data, dI/dV tunneling spectra and drifting rate values all show satisfactory results, showing the harsh vibration issue from the cryogen-free magnet system can be well solved for the simple and convenient plug-and-play probe-type STM. This becomes particularly important when resource sharing issue of the expensive cryogen-free superconducting magnet needs to be considered.

**Acknowledgments**

We sincerely thank Prof. Y.P. Sun for providing the NiSe2 single crystal. This work was supported by the National Key R&D Program of China (Nos. 2017YFA0402903 and2016YFA0401003), National Natural Science Foundation of China (Nos. 21505139, 51627901, and 11374278), National Science Foundation for Young Scientists of China (No.11504339), and Chinese Academy of Sciences Scientific Research Equipment (No. YZ201628s).